\documentclass[conference]{IEEEtran}
\usepackage{amsmath,amsfonts,amssymb, amsthm}
\usepackage{graphicx}
\usepackage{amsmath}
\usepackage{algorithm}
\usepackage{algorithmic}
\usepackage{booktabs}

\usepackage{siunitx}
\usepackage{graphicx}    % resizebox

\usepackage{tikz}
\usetikzlibrary{
  arrows.meta,   % >=latex arrow tips
  positioning    % node distance, relative placement
}

\usepackage{mathtools}   % for \coloneqq, \DeclarePairedDelimiter
\usepackage{bm}          % bold math

\newtheorem{example}{Example}
\title{Parameter Estimation of Mutual Information Maximized Channels $^{\star}$}

\author{
  \IEEEauthorblockN{Hassan Tavakoli}
  \IEEEauthorblockA{School of EECS\\
                    Oregon State University\\
                    Oregon, OR 97331, USA\\
                    tavakolh@oregonstate.edu}
  \and
  \IEEEauthorblockN{Thinh Nguyen, \emph{Senior Member, IEEE}}
  \IEEEauthorblockA{School of EECS\\
                    Oregon State University\\
                    Oregon, OR 97331, USA\\
                    thinhq@eecs.oregonstate.edu}
  \and
  \IEEEauthorblockN{Bella Bose, \emph{Life Fellow, IEEE}\thanks{ $^{\star}$ This work was supported by the National Science Foundation Grant CCF-2417898.}}
  \IEEEauthorblockA{School of EECS\\
                    Oregon State University\\
                    Oregon, OR 97331, USA\\
                    Bella.Bose@oregonstate.edu}
}

\begin{document}
\maketitle

\begingroup
\renewcommand\thefootnote{}
\footnotetext{$^{\star}$ This work was supported by the National Science Foundation under Grant No. CCF:SHF:2417898.}
\endgroup

\begin{abstract}
We study the problem of estimating a parametric discrete memoryless channel
\( p(y \mid x; \boldsymbol{\theta}) \) when the transmitter selects its input distribution
\( \pi \) to maximize mutual information under the true parameter
\( \boldsymbol{\theta}^* \).
Using only i.i.d.\ observations of the channel output, we aim to jointly
estimate the capacity-achieving input distribution \( \boldsymbol{\pi}^* \) and the true
channel parameter \( \boldsymbol{\theta}^* \).  In general, recovery of  \( \boldsymbol{\pi}^* \) and
\( \boldsymbol{\theta}^* \) can be challenging.  To that end, we propose two efficient algorithms based on the
Blahut--Arimoto (BA) optimality conditions:
(i) a bilevel fixed-point method and
(ii) an augmented Lagrangian method.
Empirical results demonstrate that both proposed algorithms successfully
recover the true \( \boldsymbol{\theta}^* \) and \( \boldsymbol{\pi}^* \), whereas a naive
maximum-likelihood approach that ignores the mutual-information
maximization constraint fails to do so.
\end{abstract}

\section{Introduction}
Estimating a parametric communication channel from noisy outputs is a classical problem with applications in calibration, coding, and sensing. Traditional methods assume that the transmitter’s input symbols or input distribution are known, so the channel can be estimated from the observed outputs and the known inputs. For example, wireless systems often use pilot sequences for this purpose. In many engineered and natural systems, however, the transmitter may jointly optimize the input distribution and the channel to maximize mutual information. This joint design is often preferred under practical constraints such as power consumption and implementation complexity.

%Estimating a parametric communication channel from noisy outputs is a classical problem with applications ranging from physical-layer calibration to adaptive coding and sensing. Traditional estimation approaches assume that the transmitter’s input symbols or input distribution are known. Under this assumption, channel parameters can be estimated using the known inputs or its distribution together with the observed outputs. For example, in wireless communication systems, a known pilot sequence is commonly transmitted, and the corresponding received signals are used to estimate channel fading coefficients. However, in many engineered systems as well as in natural processes, the transmitter actively \emph{optimizes} both the input distribution and the channel simultaneously to maximize the mutual information between input and output, i.e., to transmit the most amount of information. Specifically, in engineered systems, jointly designing the channel and the input distribution is often desirable because of resource constraints, including power consumption and the complexity of channel design.

Suppose both optimal input distribution $\boldsymbol{\pi}^*$ and channel parameter $\boldsymbol{\theta}^*$ were chosen at the transmitter in the sense that they maximize the mutual information between input and output subject to certain resource constraints. However,  $\boldsymbol{\pi}^*$ and $\boldsymbol{\theta}^*$  are unknown to the receiver. The central question in this paper is whether the receiver can accurately and efficiently recover $\boldsymbol{\pi}^*$ and $\boldsymbol{\theta}*$ based solely on an i.i.d sequence of the observed channel outputs.

Before addressing this problem, we briefly review several closely related works. Most existing settings assume that the channel parameter $\boldsymbol{\theta}$ is known,
in which case a memoryless channel is fully characterized by the conditional
probability distribution $p(y \mid x; \boldsymbol{\theta})$, where $x$ and $y$
denote the input and output symbols, respectively.
Under this assumption, the transmitter may adopt the capacity-achieving input
distribution $\boldsymbol{\pi}^*(\boldsymbol{\theta}) = \arg\max_{\boldsymbol{\pi} \in \Delta_N} I_{\boldsymbol{\pi}}(X; Y \mid \boldsymbol{\theta})$.

Often, there may be no closed-form expression for computing $ \boldsymbol{\pi}^* $; however, in general, it can be obtained using the well-known Blahut--Arimoto (BA) algorithm \cite{blahut1972,arimoto1972}.
Conversely, when the input distribution is known, but the channel parameter $\boldsymbol{\theta}$ is unknown, $\boldsymbol{\theta}$ can be estimated via the
maximum likelihood estimator as a function of $\boldsymbol{\pi}$ and the observed
outputs $y$.  When both $\boldsymbol{\pi}$ and $\boldsymbol{\theta}$ are unknown, one may treat them both as parameters and apply classical methods for estimating incomplete-data or latent-variable models, such as the Expectation--Maximization (EM) algorithm \cite{dempster1977, mclachlan2007em}. However, without explicitly imposing the maximum mutual information constraint, such approaches may fail to recover the true $\boldsymbol{\theta}^*$ and $\boldsymbol{\pi}^*$.  

Since our contribution builds on recent advances in optimization, we briefly review the most relevant literature. In particular, the machine-learning community has developed principled methods for differentiating through optimization problems and implicit layers, enabling outer-loop optimization over parameters defined by an inner optimization problem \cite{amos2017optnet,agrawal2019}. In parallel, bilevel programming has emerged as a standard framework for hyperparameter optimization and meta-learning, with a growing body of recent algorithmic and analytical developments \cite{franceschi2018}. From a broader optimization perspective, Augmented Lagrangian (AL) and related multiplier-based methods remain foundational approaches for constrained optimization, offering strong theoretical guarantees \cite{bertsekas1999}. In addition, recent work has explored mutual-information-driven parameter tuning in stochastic systems. For example, \cite{tavakoli_quantizer} studies threshold optimization in parallel stochastic quantizers by directly maximizing $I(X;Y)$, demonstrating the effectiveness of information-theoretic design. These results motivate our setting, where system parameters are also chosen to maximize mutual information, but must be inferred solely from output observations.

To this end, we propose two efficient algorithms for estimating $ \boldsymbol{\pi}^* $ and
$\boldsymbol{\theta}^*$ using only an i.i.d. sequence of observed outputs, based on the
Blahut--Arimoto (BA) optimality conditions:
(i) a bilevel fixed-point method and
(ii) an augmented Lagrangian (AL) method.
In addition, we provide some analysis for on whether or not the true $\boldsymbol{\theta}^*$ and $\boldsymbol{\pi}^*$ cannot be recovered 
accurately and efficiently (sample complexity) solely from the
observed outputs, and we present numerical results that corroborate our
theoretical findings.
\section{Preliminaries} \label{sec:preliminaries}
\subsection{Notations}
We use the following notations. Scalar parameters are denoted by italics: \(a,b,\boldsymbol{\theta}\in\mathbb{R}\). Random variables are represented by uppercase letters, e.g.\ \(X,Y\), their realizations by lowercase letters, e.g.\ \(x,y\).
Alphabets and sets are denoted calligraphic letters.
Vectors are written in boldface lowercase letters, e.g., \(\boldsymbol{\pi}=(\pi_1,\dots,\pi_N)^\top\in\mathbb{R}^N\), while matrices are written in boldface uppercase letters, e.g.\ \(\mathbf{P}\).
For a subset of indices \(S\subseteq\{1,\dots,N\}\triangleq [1:N]\) we denote by \(\boldsymbol{\pi}_S\) the subvector of \(\boldsymbol{\pi}\) with entries indexed by \(S\). Equivalently, for a vector \(\boldsymbol{x}\in\mathbb{R}^N\) the notation \(\boldsymbol{x}_S\) denotes the subvector containing \(\{x_i:i\in S\}\).
The standard simplex in \(\mathbb{R}^N\) is \(\Delta_N=\{\boldsymbol{\pi}\in\mathbb{R}^N_{\ge 0}:\ \sum_{i=1}^N\pi_i=1\}\).
We consider a discrete memoryless channel parameterized by \(\boldsymbol{\theta}\), where in general \(d=\dim(\boldsymbol{\theta})\), with transition probabilities
$W_{t|i}(\boldsymbol{\theta}) \triangleq  P_{\boldsymbol{\theta}}(y_t \mid x_i)$, for \(x_i\in\mathcal{X}\) and \(y_t\in\mathcal{Y}\).  Also, \(\partial_{\boldsymbol{\theta}}\) denotes partial differentiation with respect to \(\boldsymbol{\theta}\).  \(\nabla_x\) denotes the gradient with respect to \(x\). All \(\log\) are in base \(2\).

\subsection{Blahut--Arimoto Algorithm}
% short paragraph: existence/uniqueness remarks for BA fixed points
The Blahut-Arimoto (BA) algorithm is an iterative fixed-point algorithm that computes the channel capacity and the
capacity-achieving input distribution $ \boldsymbol{\pi}^\star $ for a given channel
$ W(\boldsymbol{\theta})$ with known $\boldsymbol{\theta}$.
Let  $ W \in \mathbb{R}^{N \times M} $) be the channel matrix of a discrete memoryless channel, then the BA fixed-point update at iteration $k$ is
\(
r(y_t) = \sum_{i=1}^N \pi_i^{(k)} W_{t|i}(\boldsymbol{\theta}), ~~~s_i = \sum_{t} W_{t|i}(\boldsymbol{\theta})
\log \frac{W_{t|i}(\boldsymbol{\theta})}{r(y_t)},
\)
followed by the update
\(
\tilde{\pi}_i^{(k)} = \pi_i^{(k)} \exp(s_i), \,
\boldsymbol{\pi}^{(k+1)} \leftarrow
b(\boldsymbol{\pi}^{(k)}, \boldsymbol{\theta})
= \frac{\tilde{\boldsymbol{\pi}}^{(k)}}{\sum_j \tilde{\pi}_j^{(k)}}.
\)
The BA algorithm iterates this mapping until convergence.
At convergence, BA reaches a \emph{fixed point} satisfying
\begin{equation}
 \label{eq:fixed}    b(\boldsymbol{\pi}, \boldsymbol{\theta}) = \boldsymbol{\pi}. 
\end{equation} This fixed point is also the Karush--Kuhn--Tucker (KKT) stationarity conditions of the convex optimization
problem
\[
\max_{\boldsymbol{\pi} \in \Delta_N} I_{\boldsymbol{\pi}}(X; Y \mid \boldsymbol{\theta}).
\]

\subsection{Augmented--Lagrangian (AL) Methods}
\label{subsec:AL_prelim}

We consider a smooth constrained maximization problem of the form
\begin{align}
\max_{u\in\mathcal{U}}\; G(u)
\qquad\text{s.t.}\qquad R(u)=0,
\end{align}
where \(G:\mathbb{R}^n\to\mathbb{R}\) and \(R:\mathbb{R}^n\to\mathbb{R}^m\) on an open set containing the feasible region, and \(\mathcal{U}\) denotes simple bounds or a convex set. The \emph{augmented Lagrangian} (AL) for maximization is defined as, \cite{hestenes1969multiplier},
\begin{align}
\mathcal{L}_{\mathrm{AL}}(u;\mu,\rho)
\triangleq  G(u) - \mu^\top R(u) - \tfrac{\rho}{2}\,\|R(u)\|_2^2,
\end{align}
with multipliers \(\mu\in\mathbb{R}^m\) and penalty \(\rho>0\).

AL algorithm is a gradient descent algorithm that alternates between (i) updating the primal variable \(u\) and (ii)  updating of the dual \(\mu\leftarrow \mu + \rho\, R(u)\).  \(\rho\) is used to ensure the residual decreases persistently. The gradient used in primal updates is
\begin{align}
\nabla_u \mathcal{L}_{\mathrm{AL}}(u;\mu,\rho)
= \nabla G(u) - \big(\mu + \rho\,R(u)\big)^\top \nabla_u R(u).
\end{align}

AL methods have been known to have numerical stability, strong convergence properties.

\section{Problem Formulation} 
\label{sec:alg}
There are three canonical settings for optimizing the mutual information between inputs and outputs of a channel:
\begin{enumerate}
\item \textbf{Fixed channel, varying input distribution:} A canonical example of this setting is the channel capacity problem. For a fixed channel $W(\boldsymbol{\theta}_0)$, the goal is to determine the input distribution $(\boldsymbol{\pi})$ that maximizes the mutual information $I_{\boldsymbol{\pi}}(X;Y \mid \boldsymbol{\theta}_0)$. In this case, $I_{\boldsymbol{\pi}}(X;Y \mid \boldsymbol{\theta}_0)$ is concave in $(\boldsymbol{\pi})$ over the probability simplex $(\Delta_N)$, leading to a convex optimization formulation that can be solved efficiently.
\item \textbf{Fixed input distribution, varying channel:} A well-known example of this setting is the privacy leakage control problem. 
In this setting, the channel parameter $\boldsymbol{\theta}$ may be interpreted as a controllable noise level or randomization mechanism that 
mediates the mapping from an input $X$ to an output $Y_{\boldsymbol{\theta}}$. A common objective is to select $\boldsymbol{\theta}$ so as to 
minimize the information revealed about $X$ the observation $Y_{\boldsymbol{\theta}}$, which can be formulated as 
$\min_{\boldsymbol{\theta} \in \Theta} I(X; Y_{\boldsymbol{\theta}})$.In general, the difficulty of this problem depends on the structural properties of the channel $W(\boldsymbol{\theta})$.
\item \textbf {Varying input, varying channel:}  This setting frequently arises in natural processes; for example, in molecular
communication channels \cite{Thomas2016}, the channel parameters may vary with
the input distribution. Moreover, in many engineered systems, such as sensing and communication platforms, channel parameters can be tuned to satisfy resource constraints, including power consumption and transmitter–receiver circuit area. As a consequence, the optimal input distribution must be jointly optimized with the channel parameters. Accordingly, we may therefore want to maximize the mutual information as
\[
I(X;Y) = \max_{\boldsymbol{\theta}, \boldsymbol{\pi}}I\big(\boldsymbol{\pi}, W(\boldsymbol{\theta})\big).
\]
\end{enumerate}

This paper considers the problem of jointly estimating the channel parameter
$\boldsymbol{\theta}$ and the input distribution $\boldsymbol{\pi}$ at the receiver.
Specifically, we adopt the following two assumptions:

\begin{enumerate}
    \item The transmitter designs the channel parameters to satisfy given resource  constraints and selects the input distribution to maximize mutual information 
\item The receiver has no knowledge of either the channel parameters or the input distribution, but observes an i.i.d. sequence of channel outputs.
\end{enumerate}
Under these assumptions, the estimation problem naturally falls into an unsupervised learning setting \cite{dempster1977em, mclachlan2008em}, in which the channel parameters are inferred without access to labeled data (i.e., the channel inputs).
 
Formally, let \(X\in\mathcal{X}\) denote the discrete channel input, $|\mathcal{X}| = N$ and let \(Y\in\mathcal{Y}\) denote the channel output, $|\mathcal{Y}| = M$. Let  $y_{1:T}=(y_1,\dots,y_T)$ denote $T$ independent observations of the output. Let the channel parameter $\boldsymbol{\theta} \in\Theta$, where $\Theta$ is due to resource constraints and the input distribution $\boldsymbol{\pi} \in \Delta_N$ (simplex probability).

For a given pair \((\boldsymbol{\theta},\boldsymbol{\pi})\) the marginal likelihood of an i.i.d sequence of observed outputs is: 
\begin{align}
\Pr(y_{1:T}\mid \boldsymbol{\theta},\boldsymbol{\pi})
=\prod_{t=1}^T \Big(\sum_{i=1}^N \pi_i\, W_{t|i}(\boldsymbol{\theta})\Big),
\end{align}
and the marginal log-likelihood is:
\begin{align}
\mathcal{L}_T(\boldsymbol{\theta},\boldsymbol{\pi})
\;=\;\sum_{t=1}^T \log\!\Big(\sum_{i=1}^N \pi_i\,W_{t|i}(\boldsymbol{\theta})\Big).
\end{align}

Since the transmitter assumes to use the input distribution that maximizes the mutual information for a given fixed channel, $\boldsymbol{\pi}$ depends on $\boldsymbol{\theta}$.  Thus, we can write $\boldsymbol{\pi}(\boldsymbol{\theta})$ to denote this dependency.  Consequently, our objective is to determine the maximum likelihood estimator $\boldsymbol{\theta}_{MLE}$:

\begin{equation}
\label{eq:MLE}
\hat{\boldsymbol{\theta}}_{MLE}=\arg\max_{\boldsymbol{\theta}} \mathcal{ L}_T\big(\boldsymbol{\theta},\boldsymbol{\pi}(\boldsymbol{\theta})\big).
\end{equation}

There are two fundamental issues associated with computing $\hat{\boldsymbol{\theta}}_{MLE}$. First, while the structure of certain channel matrices admits closed-form solutions to \eqref{eq:MLE} in special cases, such expressions are generally unavailable. One key reason is that the capacity-achieving input distribution $\boldsymbol{\pi}(\boldsymbol{\theta})$ often does not admit a closed-form characterization. Second, from the receiver’s perspective, it may not always be possible to recover the true parameter $\boldsymbol{\theta}^*$ accurately, even with a large (possibly infinite) number of samples ($T$). This naturally raises questions of identifiability and sample complexity.

To that end,  we propose two efficient algorithms to compute $\hat{\boldsymbol{\theta}}_{MLE}$: the bilevel fixed-point algorithm and the Augmented Lagrangian algorithm, both of which demonstrate strong empirical performance.  We also provide an analysis on identifiability and sample complexity in the next section.

\section{Identifiability and Variance Analysis}
In estimation theory, Fisher information matrix \cite{kay1993} is defined as:
\[
\mathcal{I}(\boldsymbol{\theta})
=
-\,\mathbb{E}_{q_{\boldsymbol{\theta}}}
\!\left[
\nabla^2_{\boldsymbol{\theta}} \log q_{\boldsymbol{\theta}}(Y)
\right]
\]
where $\nabla_{\boldsymbol{\theta}}^2$ denotes the Hessian with respect to $\boldsymbol{\theta}$, and $q_{\boldsymbol{\theta}}(y)$ is the channel output distribution.  It is typically used to quantify how good an estimator $\hat{\boldsymbol{\theta}}$.  Specifically, for any estimator, Cramer-Rao lower bound \cite{kay1993} shows  that 
\[
\mathrm{Cov}(\hat{\boldsymbol{\theta}})
\;\succeq\;
\mathcal{I}_Y(\boldsymbol{\theta})^{-1},
\]
Consequently, if $[\mathcal{I}(\boldsymbol{\theta})]_{ii}=0$, then the parameter component $\theta_i$ is not identifiable: even with an infinite number of samples, no estimator can achieve finite variance for $\theta_i$. More generally, when $[\mathcal{I}(\boldsymbol{\theta}_0)]_{ii}$ is small, accurate estimation of the $i$th component of $\boldsymbol{\theta}_0$ requires a large number of observations.

A special case arises when $q_{\boldsymbol{\theta}_1}(y)=q_{\boldsymbol{\theta}_2}(y)$ for two or more distinct parameters $\boldsymbol{\theta}_1\neq\boldsymbol{\theta}_2$, in which case the model is non-identifiable since the induced output distributions are identical.  Even more severe case is when the resulting output distribution $q_{\boldsymbol{\theta}}(y)$ does not depend on $\boldsymbol{\theta}$, or the Jacobian

\[J(\boldsymbol{\theta}) \;\triangleq \; \Big[\frac{\partial q(y;\boldsymbol{\theta})}{\partial \boldsymbol{\theta}}\Big]_{y\in\mathcal Y} = \bold{0}, \forall y \in \mathcal{Y}
\]

% \begin{theorem}[Non-Identifiability]
% \label{thm:ident-brief} Let the output distribution
% \begin{equation}
% q(y;\theta)\;=\;\sum_{i=1}^{N} \pi_i(\theta)\,W_{j|i}(\theta). \quad \forall y \in \mathcal{Y},
% \label{eq:Q}
% \end{equation}
% where $\bold{\pi}$ is the capacity-achieving input distribution.  Assume standard regularity so that \(q(y;\theta)\) is \(C^1\) in \(\theta\). Let
% \begin{equation}
% J(\theta) \;\triangleq \; \Big[\frac{\partial q(y;\theta)}{\partial \theta_j}\Big]_{y\in\mathcal Y,\; j=[1:d]}
% \end{equation}
% be the \(M\times d\) Jacobian matrix of the output distribution.
% If \[ J(\theta)  = \mathbf{0}_{M\times d} \]
% then $\theta$ is \emph{not identifiable} from outputs $Y$ alone.  In other words, no estimator based only on $y_{1:T}$ can be consistent for $\theta$.
% \end{theorem}

% \begin{proof} 
% It's straightforward to show that the condition is equivalent to the Fisher information vanishing at $\theta$.
% \end{proof}
% We showed a few identifiable and non-identifiable examples of well-known channels 

\begin{example}
Consider a binary symmetric channel BSC($\boldsymbol{\theta}$), $X, Y \in \{0, 1\}$ and the cross-over probability $\boldsymbol{\theta}$.
It is well known that the capacity-achieving input distribution for a BSC is uniform, independent of the crossover parameter $\boldsymbol{\theta}$. The resulting output distribution is also uniform, implying that all values of $\boldsymbol{\theta}$ induce statistically identical output distributions. Consequently, the parameter $\boldsymbol{\theta}$ is not identifiable.
Formally, Let $p = \Pr(X = 1)$, then
\(
q_{\theta}(1) \triangleq  \Pr(Y=1)=p(1-\theta)+(1-p)\theta,
q_{\theta}(0) \triangleq  \Pr(Y=0)= 1-q_\theta(1).
\)
For capacity-achieving input distribution $p = 0.5$, and therefore $q_\theta(1) = q_\theta(0) = p = 0.5$. 
Thus, 
\[J(\theta) \;\triangleq \; \Big[\frac{\partial q(y;\theta)}{\partial \theta}\Big]_{y\in\mathcal Y} = 0.
\]
Therefore, $\theta$ is non-identifiable.
\end{example}

{\em Remark:} We note that if the transmitter does not use the capacity-achieving input to send information, then $\theta$ can be estimated. 

\begin{example}
 Consider a binary erasure channel (BEC ($\theta)$) where \(\Pr(X=1)=p\), and \(P(Y=?\mid X=0)=P(Y=?\mid X=1)=\theta\), \(\theta\in [0,1]\), \(X \in \{0,1\}\).
 The output distribution is:
\(
r_1 \triangleq q_\theta(1) \triangleq  \Pr(Y=1)=p(1-\theta), 
r_0 \triangleq q_\theta(0) \triangleq  \Pr(Y=0)=(1-p)(1-\theta),
r_? \triangleq q_\theta(?) \triangleq  \Pr(Y=?)=\theta.
\)
Let \(S_1\), \(S_0\) and \(S_?\) be the number of outputs "1", "0", and "?", respectively.   The marginal log-likelihood can be written as 
\(
   \mathcal{L}_T(\boldsymbol{\pi},\theta)
= S_1\log\big(r_1\big) + S_0\log\big(r_0\big)+ S_?\log\big(r_?\big), 
\)
$T = S_1 +  S_0 + S_?$.
For any BEC, the capacity-achieving input is uniform \cite{CoverThomas2006}, i.e., $p = \frac{1}{2}$. Set $\partial_{\theta} \mathcal{L_T} = 0$ to find $\hat{\theta}_{MLE}$,
\(   \frac{\partial_{\theta} \mathcal{L}}{\partial\theta}
= -\frac{S_1+S_0}{1-\theta}+\frac{S_?}{\theta}=0
\)
which gives unique
\(
\hat{\theta}=\frac{S_?}{T},
\)
and the corresponding channel capacity is \(C=1-\frac{S_?}{T}\).
So $\theta$ is identifiable.  
 \end{example}
To analyze the variance of $\hat{\theta}_{MLE}$, we compute the Fisher information using $p= 0.5$ as 
\begin{align*}
\mathcal{I}(\theta) &=
\sum_{y\in\mathcal{Y}}
q_\theta(y)
\left(
\frac{\partial}{\partial\theta}\log q_\theta(y)
\right)^2  
	    = \frac{1}{\theta(1-\theta)}
\end{align*}
{\em Remark:}
The Fisher information diverges as $\theta \to 0$ or $\theta \to 1$, reflecting
the fact that the channel becomes nearly deterministic in these regimes.
For $T$ i.i.d.\ channel uses, the variance of $\hat{\theta}_{MLE}$ scales as
$\frac{\theta(1-\theta)}{T}$.

\section{Algorithms}
\label{sec:algorithms}
Before discussing the two proposed algorithm to determine $\hat{\boldsymbol{\theta}}_{MLE}$, we briefly discuss why imposing maximum mutual information will generally help algorithms converge quicker to the correct $\boldsymbol{\theta}$.

\subsection{MLE Without Mutual Information Maximization}
Suppose we jointly estimate $\boldsymbol{\theta},\boldsymbol{\pi}$ by maximum likelihood. Without additional constraints, the problem may be non-identifiable, since different channel–input pairs can induce the same output distribution. The example below illustrates this ambiguity.

 %Suppose we attempt to jointly estimate $(\boldsymbol{\theta},\boldsymbol{\pi})$ via maximum likelihood. To ensure identifiability and obtain a unique solution, it is necessary to restrict the feasible set, for example by imposing structural constraints, fixing components of $\boldsymbol{\pi}$, or introducing informative priors on $\boldsymbol{\theta}$. Absent such restrictions, distinct channel–input pairs may be observationally indistinguishable. The example below illustrates that two different channel matrices, paired with different input distributions, can induce the same marginal distribution on the output $Y$.

Let \(W(\boldsymbol{\theta})\in\mathbb{R}^{N\times M}\) denote the channel matrix whose \(i\)-th row is the conditional distribution \(P_{\boldsymbol{\theta}}(Y\!\mid\! X=x_i)\). Suppose there exist \(\boldsymbol{\theta}_1\neq\boldsymbol{\theta}_2\) and \(\varepsilon>0\) such that
\begin{equation}
\|W(\boldsymbol{\theta}_1)-W(\boldsymbol{\theta}_2)\|_\infty \;>\; \varepsilon,
\label{eq:Wdiff}
\end{equation}
but there also exist different input distributions \(\boldsymbol{\pi}^{(1)},\boldsymbol{\pi}^{(2)}\in\Delta_N\) with
\begin{equation}
\boldsymbol{\pi}^{(1)}W(\boldsymbol{\theta}_1) \;=\; \boldsymbol{\pi}^{(2)}W(\boldsymbol{\theta}_2),
\label{eq:marginal_equal}
\end{equation}
where \(\|\cdot\|_\infty\) means the entrywise sup-norm. Then Eq. \eqref{eq:marginal_equal} implies the two parameter pairs \((\boldsymbol{\theta}_1,\boldsymbol{\pi}^{(1)})\) and \((\boldsymbol{\theta}_2,\boldsymbol{\pi}^{(2)})\) induce the identical marginal distribution for \(Y\). Therefore, for any observed data \(y_{1:T}\) the marginal log-likelihoods satisfy
\begin{align}
\mathcal L_T(\boldsymbol{\theta}_1,\boldsymbol{\pi}^{(1)}) \;=\; \mathcal L_T(\boldsymbol{\theta}_2,\boldsymbol{\pi}^{(2)}),
\end{align}
and
\(\displaystyle \max_{\boldsymbol{\theta},\boldsymbol{\pi}}\mathcal L_T(\boldsymbol{\theta},\boldsymbol{\pi})\)
cannot distinguish \(\boldsymbol{\theta}_1\) from \(\boldsymbol{\theta}_2\) using \(Y\) alone (unless extra constraints or priors are imposed).
\begin{example} \label{ex:MLE}
Let \(W\) have a 2-input symbol and a 3-output channel. Its probability values at two channel parameters of \(\boldsymbol{\theta}_1\) and \(\boldsymbol{\theta}_2\) are:
\[
W(\boldsymbol{\theta}_1)=
\begin{bmatrix}
0.8 & 0.1 & 0.1\\[4pt]
0.1 & 0.8 & 0.1
\end{bmatrix},\qquad
W(\boldsymbol{\theta}_2)=
\begin{bmatrix}
0.5 & 0.4 & 0.1\\[4pt]
0.3 & 0.6 & 0.1
\end{bmatrix}.
\]
\end{example}
Note that \(\|W(\boldsymbol{\theta}_1)-W(\boldsymbol{\theta}_2)\|_\infty \ge 0.3\); choose \(\varepsilon=0.1\) to satisfy Eq. \eqref{eq:Wdiff} and define input distributions
\(
\boldsymbol{\pi}^{(1)}=\begin{bmatrix} \tfrac{3}{7} \,\, \tfrac{4}{7} \end{bmatrix},\)
and
\(
\boldsymbol{\pi}^{(2)}=\begin{bmatrix} \tfrac{1}{2} \,\, \tfrac{1}{2} \end{bmatrix}.
\)
Compute the induced output marginal for each pair, Eq. \eqref{eq:marginal_equal} will be satisfied and equal to \([0.4 \,\, 0.5 \, \, 0.1]\). 

Next, we describe two algorithms that employ the constraint on input distribution $\boldsymbol{\pi}(\boldsymbol{\theta})$ that maximizes the mutual information.

\subsection{Bilevel Fixed Point Algorithm}
Bilevel optimization refers to a class of optimization problems in which one problem is nested inside another. Specifically, an upper-level problem’s objective depends on the solution of a lower-level problem. Formally, it can be written as
\[
\max_{\boldsymbol{\theta} \in \Theta} G(\boldsymbol{\theta}, \boldsymbol{\pi}) \quad \text{subject to} \quad \boldsymbol{\pi} \in \arg\max_{\boldsymbol{\pi} \in \Delta (\boldsymbol{\theta})} F(\boldsymbol{\theta}, \boldsymbol{\pi}),
\]
 where $G(\boldsymbol{\theta},\boldsymbol{\pi})$ and $F(\boldsymbol{\theta},\boldsymbol{\pi})$ are the upper-level and lower-level objectives, respectively, and $\Delta(\boldsymbol{\theta})$ represents constraints for the lower-level problem.
A common approach to solving a bilevel optimization problem is to iteratively alternate between optimizing the lower-level objective and the upper-level objective, repeating this process until convergence is achieved. Typically, gradient algorithms are used to optimize for each of the objective, thus the respective gradients need to be computed. 

 In our problem, the objective function log-likelihood $\mathcal{L}_T(\boldsymbol{\theta}, \boldsymbol{\pi})$ represents the upper-level objective $G(\boldsymbol{\theta}, \boldsymbol{\pi})$, and the mutual information can be thought of as the lower-level objective $F(\boldsymbol{\theta}, \boldsymbol{\pi})$.  However, since the $\boldsymbol{\pi}^*$ that maximizes the mutual information for a fixed $\boldsymbol{\theta}$ must satisfy the BA fixed point condition in \ref{eq:fixed}, i.e., $\boldsymbol{\pi}(\boldsymbol{\theta})$ depends on $\boldsymbol{\theta}$, we can compute the gradient of $\mathcal{L}_T(\boldsymbol{\theta}, \boldsymbol{\pi}(\boldsymbol{\theta}))$ as:
\begin{equation}
\label{eq:fixed_point_grad}
\nabla_{\boldsymbol{\theta}} \mathcal{L}(\boldsymbol{\theta})
	=
	\nabla_{\boldsymbol{\theta}} \mathcal{L}(\boldsymbol{\theta},\boldsymbol{\pi})
	+
	\big(\nabla_{\boldsymbol{\pi}} \mathcal{L}(\boldsymbol{\theta},\boldsymbol{\pi})\big)
	\,\nabla_{\boldsymbol{\theta}} \boldsymbol{\pi}(\boldsymbol{\theta}),
\end{equation}

To compute $\nabla_{\boldsymbol{\theta}} \boldsymbol{\pi}(\boldsymbol{\theta})$, we use the BA fixed point optimality condition for maximizing the mutual information and define
$$R(\boldsymbol{\theta},\boldsymbol{\pi})=b(\boldsymbol{\theta},\boldsymbol{\pi}) - \boldsymbol{\pi} = \boldsymbol{0},$$
where \(R:\mathbb{R}^n\times\mathbb{R}^d\to\mathbb{R}^n\).
Differentiating $R(\boldsymbol{\theta},\boldsymbol{\pi}(\boldsymbol{\theta}))=\boldsymbol{0}$ with respect
to $\boldsymbol{\theta}$ yields
\[
\nabla_{\boldsymbol{\theta}} R(\theta,\boldsymbol{\pi})
+
\nabla_{\boldsymbol{\pi}} R(\boldsymbol{\theta},\boldsymbol{\pi})
\,\nabla_{\boldsymbol{\theta}} \boldsymbol{\pi}(\boldsymbol{\theta})
=
\boldsymbol{0}.
\]
Since $R(\boldsymbol{\theta},\boldsymbol{\pi})=b(\boldsymbol{\theta},\boldsymbol{\pi})-\boldsymbol{\pi}$,
we have
\[
\nabla_{\boldsymbol{\theta}} R(\boldsymbol{\theta},\boldsymbol{\pi})
=
\nabla_{\boldsymbol{\theta}} b(\boldsymbol{\theta},\boldsymbol{\pi}),
\qquad
\nabla_{\boldsymbol{\pi}} R(\boldsymbol{\theta},\boldsymbol{\pi})
=
\nabla_{\boldsymbol{\pi}} b(\boldsymbol{\theta},\boldsymbol{\pi}) - I_n,
\]
where $I_n$ is the $n\times n$ identity matrix.
Assuming that $I_n-\nabla_{\boldsymbol{\pi}} b(\boldsymbol{\theta},\boldsymbol{\pi})$ is
invertible, we solve for $\nabla_{\boldsymbol{\theta}} \boldsymbol{\pi}(\boldsymbol{\theta})$:
\begin{equation}
\label{eq:fixed_solver}
	\nabla_{\boldsymbol{\theta}} \boldsymbol{\pi}(\boldsymbol{\theta})
	=
	\big(I_n-\nabla_{\boldsymbol{\pi}} b(\boldsymbol{\theta},\boldsymbol{\pi})\big)^{-1}
	\nabla_{\boldsymbol{\theta}} b(\boldsymbol{\theta},\boldsymbol{\pi}).
\end{equation}
Using Eqs. \ref{eq:fixed_point_grad} and \ref{eq:fixed_solver}, the proposed bilevel fixed point Algorithm \ref{alg:profile-short} employs gradient step until the outer loop converges.

\begin{algorithm}[ht]
\caption{Bilevel Fixed Point Algorithm}\label{alg:profile-short}
\begin{algorithmic}[1]
\REQUIRE Data $y_{1:T}$, init $\boldsymbol{\theta}^0,\boldsymbol{\pi}^0$, BA tol $\tau_{\rm BA}$, outer iters $K$, step sizes $\{\eta_k\}$
\FOR{$k=0,\dots,K-1$}
  \STATE $\boldsymbol{\pi}\leftarrow\textsc{BA}(\boldsymbol{\pi},\boldsymbol{\theta};\ \text{tol}=\tau_{\rm BA})$ 
  \STATE $\nabla_{\boldsymbol{\theta}} \boldsymbol{\pi}(\boldsymbol{\theta})
	\leftarrow
	\big(I_n-\nabla_{\boldsymbol{\pi}} b(\boldsymbol{\theta},\boldsymbol{\pi})\big)^{-1}
	\nabla_{\boldsymbol{\theta}} b(\boldsymbol{\theta},\boldsymbol{\pi})$
  \STATE $\nabla_{\boldsymbol{\theta}} \mathcal{L}(\boldsymbol{\theta})
	\leftarrow
	\nabla_{\boldsymbol{\theta}} \mathcal{L}(\boldsymbol{\theta},\boldsymbol{\pi})
	+
	\big(\nabla_{\boldsymbol{\pi}} \mathcal{L}(\boldsymbol{\theta},\boldsymbol{\pi})\big)
	\,\nabla_{\boldsymbol{\theta}} \boldsymbol{\pi}(\boldsymbol{\theta})$
  \STATE $\boldsymbol{\theta}\leftarrow\boldsymbol{\theta}+\eta_k\,\nabla_{\boldsymbol{\theta}} \mathcal L$
  \STATE \textbf{stop} if outer convergence
\ENDFOR
\RETURN $(\boldsymbol{\theta},\boldsymbol{\pi})$
\end{algorithmic}
\end{algorithm}
% Key: inner BA is solved to (near) convergence each outer step.
\begin{algorithm}[ht]
\caption{Augmented Lagrangian Algorithm}\label{alg:al-short}
\begin{algorithmic}[1]
\REQUIRE Data $y_{1:T}$, init $(\boldsymbol{\theta}^0,\boldsymbol{\pi}^0,\mu^0,\rho^0)$, outer iters $K$, BA inner steps $k_{\rm in}$, step sizes $\{\eta_k\}$
\FOR{$k=0,\dots,K-1$}
   \STATE for $t=1,\dots,k_{\rm in}$ do $\;\boldsymbol{\pi}\leftarrow b(\boldsymbol{\pi},\boldsymbol{\theta})$
  \STATE $R\leftarrow b(\boldsymbol{\pi},\boldsymbol{\theta})-\boldsymbol{\pi}$
  \STATE $\nabla_{\boldsymbol{\theta}} L_{\mathrm{AL}}\leftarrow \nabla_{\boldsymbol{\theta}} \mathcal L_T - (\mu+\rho R)^\top \nabla_{\boldsymbol{\theta}} R$
  \STATE $\boldsymbol{\theta}\leftarrow\boldsymbol{\theta}+\eta_k\,\nabla_{\boldsymbol{\theta}} L_{\mathrm{AL}}$
  \STATE \textbf{stop} if $\|R\|$ and $\|\boldsymbol{\theta}\!\!-\!\!\boldsymbol{\theta}_{\text{prev}}\|$ small
\ENDFOR
\RETURN $(\boldsymbol{\theta},\boldsymbol{\pi},\mu)$
\end{algorithmic}
\end{algorithm}
\subsection{Augmented Lagrangian Algorithm}
In this proposed algorithm, the BA optimality condition is not strictly adhered to.  Rather, we add a penalty term for the BA optimality condition when it does not hold. The algorithm is based on standard update steps of the primal and dual variables as described in Section \ref{subsec:AL_prelim}.
The main advantage of AL is that it uses only a few inner BA iterations and avoids the implicit-differentiation step required by the bilevel method. As a result, it has lower per-iteration cost and better scalability with \(N\). Its improved empirical accuracy is observed in our experiments, but this is not claimed as a theoretical guarantee.

\section{Simulation results}
\label{sec:sim}
We evaluate the AL and bilevel algorithms alongside Joint-ML, a standard method that treats both $\boldsymbol{\pi}$ and $\boldsymbol{\theta}$ as free variables, without enforcing the maximum mutual information constraint, on a discrete memoryless Gaussian-like channel. We assume $N=10$, $M=50$; $T=200{,}000$; $k_{in}=6$; BA tolerance $\tau_{\mathrm{BA}}=10^{-10}$ (max BA iters $=2{,}000$); Adam learning rate $=10^{-2}$; true parameter $\theta^\star=0.7$; search range $(\theta_{\min},\theta_{\max})=(0.1,5)$.
\paragraph{Channel model} The following discrete channel used in simulations
\begin{equation}
\label{eq:pyx}
W_{j\mid i}(\boldsymbol{\theta})\;=\;\frac{\exp\!\big(- \frac{(y_j - x_i)^2}{\theta}\big)}
{\sum_{k=1}^M \exp\!\big(- (y_k - x_i)^2 / \theta\big)},
\end{equation}
where \(i=[1:N],j=[1:M]\), and also we assume that $\theta_{\min}\le\theta\le\theta_{\max}$.

\begin{figure}[ht]
  \centering
  \includegraphics[width=0.5\linewidth]{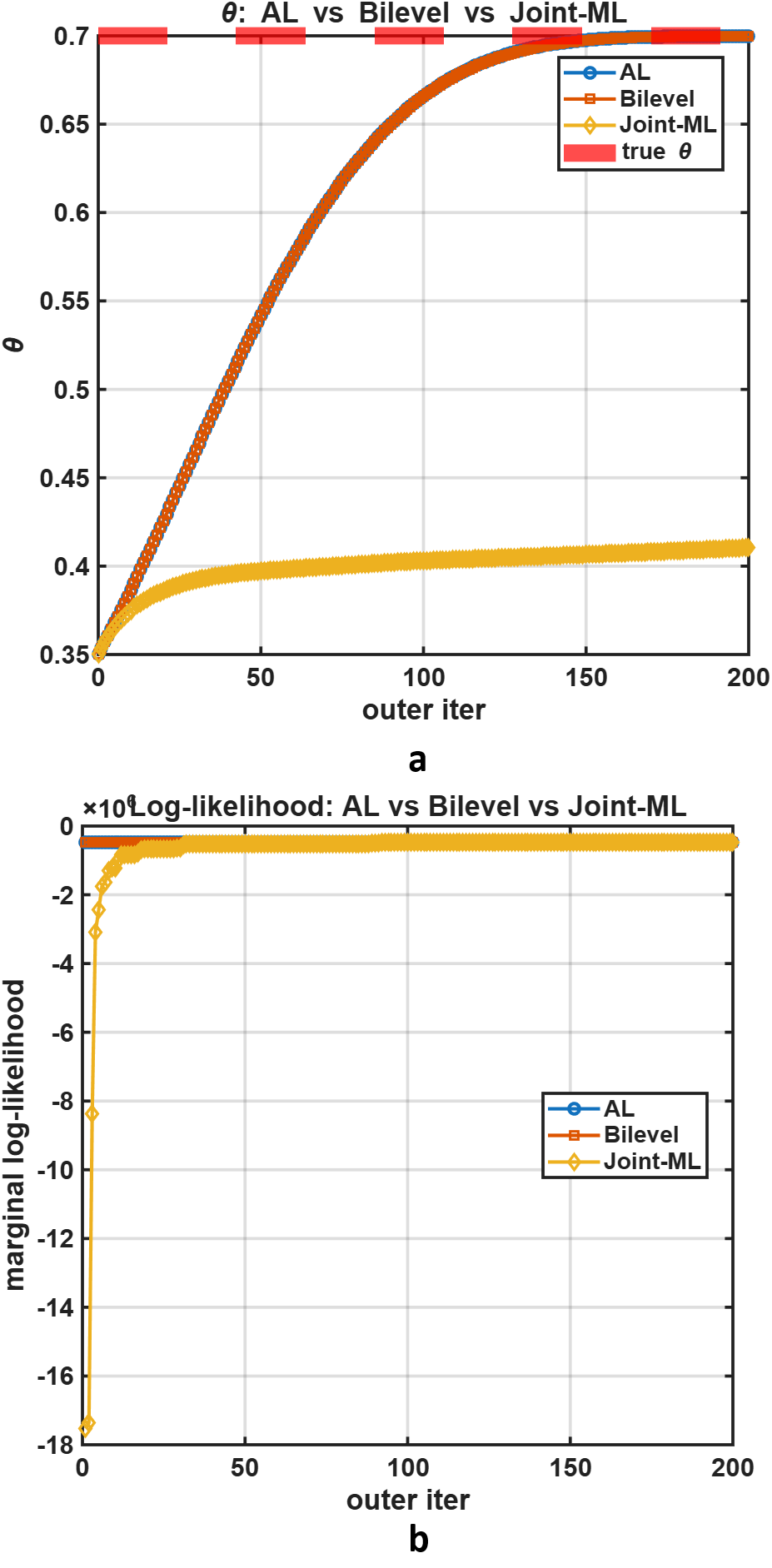}
\caption{Comparison across outer iterations. (a) Estimated \(\theta^{(k)}\): AL and bilevel converge to the true \(\theta^\star\) (dashed), while Joint-ML converges to a wrong value. (b) Log-likelihood \(\mathcal L_T\): AL and bilevel attain essentially identical values, whereas MLE reaches same likelihood at a different point, Example~\eqref{ex:MLE}.}
  \label{fig:sim_summary1}
\end{figure}

\begin{figure}[ht]
  \centering
  \includegraphics[width=0.5\linewidth]{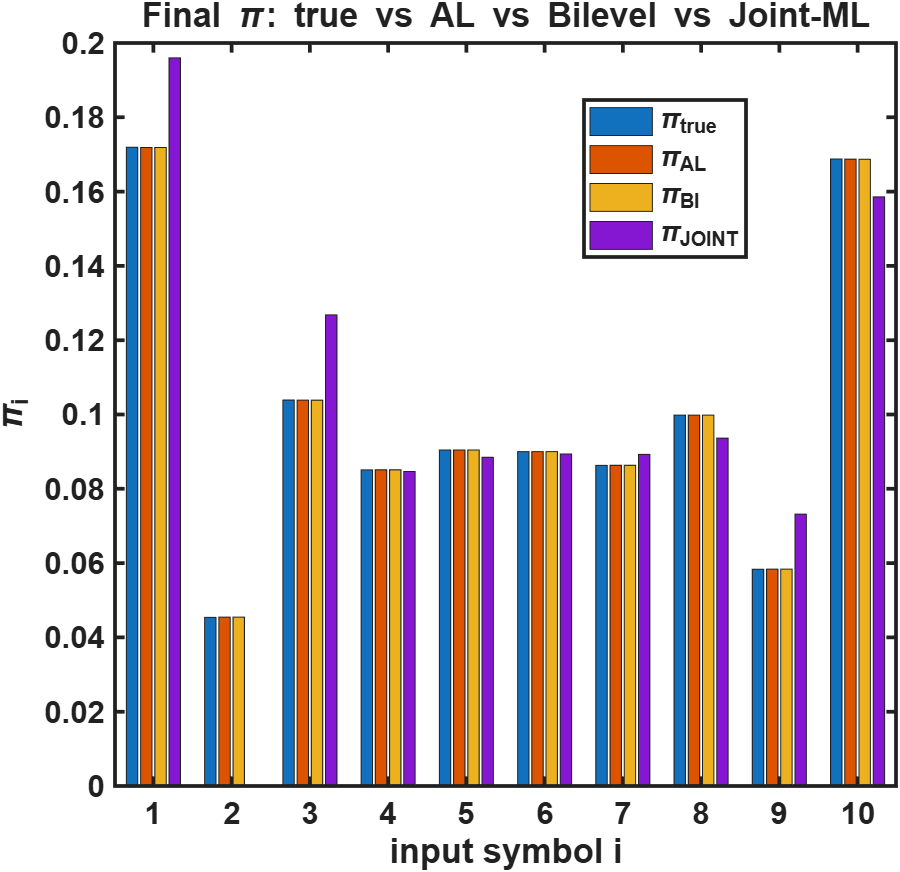}
\caption{Final input distribution recovery relative to the true capacity-achieving \(\boldsymbol{\pi}\). AL and Bilevel recover \(\boldsymbol{\pi}\) to high precision (differences on the order of \(10^{-1}\)) with KL divergences \(9.0\times10^{-8}\) (AL) and \(1.05\times10^{-7}\) (Bilevel); Joint-ML converges to an incorrect solution (max difference \(\approx0.1296\), KL \(=1.0743\)).}
  \label{fig:sim_summary2}
\end{figure}

\paragraph{Convergence behavior}
We compare the AL, bilevel, and Joint-ML algorithms on the proposed channel. For each run we record and plot, the outer \(\theta\)-trajectory, marginal log-likelihood, BA residual \(\|R\|_1\) and the estimated input distribution \(\boldsymbol{\pi}\) at termination. We also report wall-clock time and final absolute error \(|\hat\theta-\theta_{\rm true}|\). Example figures generated by the script are referenced below. Convergence of AL and bilevel to the same true parameter and marginal likelihood is shown in Fig.~\ref{fig:sim_summary1}, while comparable residuals and input-distribution accuracy achieved by AL with fewer expensive BA solves, Fig.~\ref{fig:sim_summary2}.
Figures~\ref{fig:sim_summary1} and~\ref{fig:sim_summary2} also show that a standard MLE algoritm can converge to a wrong parameter \((\theta_{ML},\pi_{ML})\) with high likelihood far from the true ones.

\paragraph{Computational efficiency}
We compare the merged AL scheme and the Bilevel scheme over 8 independent trials. We ran each method for 8 trials on a PC with an Intel(R) Core(TM) i5-4570S CPU (2.90\,GHz).
Although both methods converge to nearly identical final estimates (median $|\hat\theta-\theta^\star|$: AL $=5.57\times10^{-4}$, Bilevel $=6.10\times10^{-4}$), AL requires far fewer expensive Blahut--Arimoto (BA) solves: median $33{,}546$ full BA solves and $4.74\,$s wall-clock time for AL vs.\ $50{,}275$ solves and $5.13\,$s for Bilevel (a reduction of $33.3\%$ in BA solves).
This translates into a median runtime reduction of 7.7\% (4.74\,s vs.\ 5.13\,s) while preserving, and slightly improving, estimation accuracy. It demonstrates that AL attains a target accuracy with substantially fewer BA iterations.

\section{Conclusion}
\label{sec:conclusion}

We proposed a merged AL solver for estimating channel parameters when the transmitter selects its input via BA. By treating the BA fixed-point, the method alternates a few cheap damped BA-map steps, multiplier updates, and outer \(\theta\)-updates using a hybrid gradient strategy. In experiments the AL scheme matched the bilevel accuracy while requiring substantially fewer expensive full BA solves and modestly less runtime, demonstrating a clear computational advantage in practice.  The method is simple to implement and compatible. Limitations include the need for local BA fixed-point uniqueness and some tuning of penalty updates.

\bibliographystyle{IEEEtran}
\bibliography{bibU}

\end{document}